\documentstyle[aps,preprint,axodraw]{revtex}
\newcommand{\beq}{\begin{equation}}
\newcommand{\eeq}{\end{equation}}
\newcommand{\beqa}{\begin{eqnarray}}
\newcommand{\eeqa}{\end{eqnarray}}
\newcommand{\beqan}{\begin{eqnarray*}}
\newcommand{\eeqan}{\end{eqnarray*}}

\tighten
\begin{document}

\title{\rightline{\small SINP/TNP/97-24}
\rightline{\small SBNC/97-12-1}
{\bf Lattice Fermions and the Chiral Anomaly}}

\author{{\bf H. Banerjee$^a$ and Asit K. De$^b$} \\
$^a${\sl S.N.Bose National Centre for Basic Sciences,
JD Block, Salt Lake, Calcutta 700 091, India\\
e-mail:banerjee@boson.bose.res.in\\} 
$^b${\sl Theory Group, Saha Institute of Nuclear Physics,
1/AF Salt Lake, Calcutta 700 064, India\\
e-mail:de@tnp.saha.ernet.in}}

\date{December,1997}


\maketitle

%


\begin{abstract}
We show in the Wilson model that the contribution of the regular mass term
to the four-divergence of the axial vector current in weak coupling
perturbation theory is not zero in the chiral limit and is precisely the
axial anomaly. Explicit breaking of chiral symmetry in the Wilson term is
not relevant for the result. The ABJ anomaly is generated by the fermion
mass term also with a chirally symmetric irrelevant term.
\end{abstract}

\bigskip

\noindent
{\bf Introduction}
The proliferation of species known as {\it species doubling} is a major
problem in formulating a lattice action for fermions. 
On a \hbox{$3+1$} dimensional
hypercubic euclidean lattice, there are in the spectrum 15 unwanted doublers
in addition to the physical fermion. The doublers can be removed from the
spectrum in the continuum limit (i.e., as the lattice spacing $a\rightarrow
0$), but only at a price. This, in essence, is the message of the no-go
theorem of Nielsen and Ninomiya \cite{NN81}. The price to be paid is to
necessarily abandon one or more of the properties of (i) locality, 
(ii) hypercubic
symmetry, (iii) reflection positivity, and (iv) chiral symmetry, in the lattice
action for fermions \cite{Pelli87}.   

The most popular model for lattice fermions, the Wilson model
\cite{Wilson77}, solves the problem of doubling through an {\it irrelevant
term}, the Wilson term, which breaks chiral symmetry explicitly. This
specific choice, {\it viz.}, breaking chiral symmetry, turns out to have a more
profound reason. It is  
generally accepted that in order to obtain the
Adler-Bell-Jackiew (ABJ) anomaly in perturbation theory in the continuum
limit, the irrelevant term in the lattice fermion action, needed to remove
the spurious fermion doublers, should break chiral symmetry explicitly.
Indeed, the contribution of the Wilson term to the four-divergence of the
axial current is treated as the driving term for the ABJ anomaly
\cite{KaSm81,Kerler83}. This has been a major hurdle for lattice formulation
of chiral gauge theories, because the regulator breaks the gauge symmetry
explicitly.

We thus arrive at an impasse. Whereas, a lattice regulator preserving
chiral symmetry is desirable for construction of a chiral gauge theory, it
appears that precisely this symmetry needs to be broken explicitly in the 
irrelevant term of the lattice fermion action in order to produce the
ABJ anomaly in the continuum limit. For a way out of this apparent impasse
it is instructive to look at the pioneering work of Lee and Nauenberg (LN)
\cite{LN64}. LN showed that helicity-flip interactions in QED, while
forbidden for strictly massless electrons
because of $\gamma_5$-invariance, survives in the chiral limit if one
started with QED with massive electrons. The result of LN, therefore,
suggests that the fermion mass term has more a dynamical role than just soft
symmetry breaking. The clue was indeed picked up by Dolgov and Zakharov
\cite{DZ71} who showed that the mass of a Dirac fermion could generate
the ABJ anomaly in the axial current. The phenomenon is strongly reminiscent
of ferromagnetism, which is triggered by a weak magnetic field but survives
even after the latter is switched off.

In the following, we show that also on lattice the mass of a fermion
generates the ABJ anomaly as in \cite{LN64,DZ71}. In the Wilson model, the
continuum limit of the contribution of fermion mass $m$ to the
four-divergence of the axial current does not vanish in the chiral limit
$m\rightarrow 0$, and, what is most interesting, coincides with the ABJ
anomaly. The analysis needs rather mild constraints on the irrelevant term.
Breaking chiral symmetry, for instance, is not required. Indeed, as we
demonstrate in the following, the irrelevant term may just as well be chosen
as chirally symmetric, and one can still generate the ABJ anomaly from the
fermion mass term. 
\vskip 0.5cm

\noindent
{\bf ABJ anomaly in Wilson model.}
In the following, we work in lattice QED with Wilson fermions. Contribution
from the fermion mass to the four-divergence of the axial current in weak
coupling perturbation in the second order is given by the amplitudes of
diagrams (a), (b) and (c) in Fig.1.

\begin{center}
\begin{picture}(500,200)(0,0)
\Vertex(80,175){2.5}
\Vertex(250,175){2.5}
\Vertex(420,175){2.5}
\Line(80,175)(50,125)
\Line(250,175)(220,125)
\Line(80,175)(110,125)
\Line(250,175)(280,125)
\ArrowLine(110,125)(50,125)
\ArrowLine(280,125)(220,125)
\Oval(420,145)(30,20)(0)
\ArrowLine(400,144)(400,146)
\Photon(50,125)(50,85){-2}{4}
\Photon(110,125)(110,85){2}{4}
\Photon(220,125)(220,85){-2}{4}
\Photon(280,125)(280,85){2}{4}
\Photon(420,115)(390,85){1.6}{4}
\Photon(420,115)(450,85){-1.6}{4}
\Text(80,185)[]{$2im\gamma_5$}
\Text(250,185)[]{$2im\gamma_5$}
\Text(420,185)[]{$2im\gamma_5$}
\Text(42,105)[]{$p$}
\Text(118,105)[]{$q$}
\Text(212,105)[]{$q$}
\Text(289,105)[]{$p$}
\Text(397,100)[]{$p$}
\Text(448,100)[]{$q$}
\Text(50,77)[]{$\mu$}
\Text(110,77)[]{$\nu$}
\Text(220,77)[]{$\nu$}
\Text(280,77)[]{$\mu$}
\Text(390,79)[]{$\mu$}
\Text(450,79)[]{$\nu$}
\Text(80,117)[]{$l$}
\Text(250,117)[]{$l$}
\Text(395,145)[]{$l$}
\Text(80,50)[]{(a)}
\Text(250,50)[]{(b)}
\Text(420,50)[]{(c)}
\Text(250,10)[]{Fig.1}
\end{picture}
\end{center}
 
The external vertex in each diagram is $2im
\gamma_5$. In the Wilson model, amplitude (c) vanishes, whereas, amplitudes
(a) and (b) are equal and gauge invariant. The amplitudes (a) and (b) may,
therefore, be combined and expressed in the notation used by Karsten and
Smit \cite{KaSm81} as
\beq
D^{(a+b)}_{\mu\nu} = -4g^2 m 
                     \int_l Tr [i\gamma_5 S(l-p) V_\mu(l-p,l) S(l)
                     V_\nu(l,l+q) S(l+q)],    
\eeq
where the range of integration of the loop momentum $l$ is the Brillouin
zone ($-\frac{\pi}{a} \leq l_\alpha \leq \frac{\pi}{a}$). The fermion
propagator $S(p)$ and the one-photon vertex $V_\mu(p,q)$ are
\beqa
S(p) &=& \left[ \sum_\mu \gamma_\mu \frac{\sin ap_\mu}{a} + 
         \frac{M_c(ap)}{a} +m \right]^{-1}, \\
V_\mu &=& \gamma_\mu \cos \frac{a}{2}(p+q)_\mu + r \sin \frac{a}{2}(p+q)_\mu.
\eeqa
The Wilson term ($M_c/a$) which removes the doublers is given by
\beq
\frac{M_c(ap)}{a} = \frac{r}{a} \sum_\mu (1-\cos ap_\mu).
\eeq

Gauge invariance dictates the tensor structure
\beq
D^{(a+b)}_{\mu\nu} \propto \epsilon_{\mu\nu\alpha\beta} p_\alpha q_\beta .
\eeq
Reisz power counting \cite{Reisz,Luescher}, then gives for the lattice
Feynman integral (1) an {\it effective} degree (see discussion below)
\beq
deg\;\; D^{(a+b)}_{\mu\nu} = -2.
\eeq
The continuum limit of the lattice integral (1) is, therefore, given,
according to the Reisz theorem \cite{Reisz,Luescher}, by the integral of
continuum limit of the integrand
\beqa
\lim_{a\rightarrow 0} D^{(a+b)}_{\mu\nu} &=& ig^2 
     \epsilon_{\mu\nu\alpha\beta}p_\alpha q_\beta
     16 \int^\infty_{-\infty} \frac{d^4 l}{(2\pi)^4} \frac{m^2}{(l^2+m^2)^3}
     \nonumber \\
&=& \frac{i}{2\pi^2} g^2 \epsilon_{\mu\nu\alpha\beta} p_\alpha q_\beta .
\eeqa

It should be noted that the leading behaviour, according to the Reisz power
counting, of the lattice Feynman integral (1) arises from the piece
contributed by Dirac trace in the numerator which depends linearly on the
Wilson term $M_c/a$. The degree of this term is $ 1$.
Two powers of the lattice spacing $a$, one each with $p_\alpha$ and
$q_\beta$ brings down the degree to $-1$.
However, in the continuum limit the leading term vanishes. It is in this
sense that the {\it effective} degree of (1) is $-2$ as stated in (6).

It is remarkable that the Wilson term (4) also generates a contribution
equal in magnitude to (7) but opposite in sign. To see this, it is
convenient to combine the usual mass term with the Wilson term
\beq
\frac{M(ap)}{a} = \frac{1}{a}\left[am+r\sum_\mu(1-\cos ap_\mu a)\right]
\eeq
and consider the amplitude for the triangle diagrams (a) and (b) with this
generalized momentum dependent mass at the vertex,
\beqa
&&D_{\mu\nu}^M= \nonumber \\
&&-2g^2\int_l\frac{1}{a}\{M(al+aq)+M(al-ap)\}
              \;\;Tr\left[i\gamma_5 S(l-p)V_\mu(l-p,l)S(l)V_\nu(l,l+q)
              S(l+q)\right].
\eeqa
The leading term of the gauge invariant form of the lattice Feynman integral
thus obtained
\beq
D_{\mu\nu}^M=ig^2\epsilon_{\mu\nu\alpha\beta}p_\alpha q_\beta \;\;16 \;\;
             \int_l \frac{\left[M^2(l)-rM(l)\sum_\lambda
(\frac{\sin^2 l_\lambda}{\cos l_\lambda})\right]
\prod_\lambda \cos l_\lambda}{\left[M^2(l)+\sum_\lambda\sin^2 l_\lambda
\right]^3}
\eeq
has Reisz degree zero, and, therefore, is not amenable to the Reisz theorem.
Thanks, however, to the trigonometric identity \cite{KaSm81}
\beqa
-rM(l)\sin^2 l_\lambda &=& \cos l_\lambda \left[\sin^2 l_\lambda - \frac{1}{4}
       (M^2(l)+\sum_\lambda\sin^2 l_\lambda)\right] \nonumber \\
& & +\frac{1}{4}(M^2(l)+\sum_\lambda\sin^2 l_\lambda)^3
    \frac{\partial}{\partial l_\lambda}\left[\frac{\sin l_\lambda}
    {(M^2(l)+\sum_\lambda\sin^2 l_\lambda)^2}\right],
\eeqa
the lattice Feynman integral (10) is easily seen to vanish
\beq
D_{\mu\nu}^M = 0.
\eeq

Combined with (7), the result (12) suggests that the {\em anomalies} arising
from the 15 doublers cancel exactly the anomaly from the physical fermion 
\cite{Fuji}. In order that the result (7) translates into the familiar form
\beq
\langle \partial_\mu J_{\mu 5}(x)\rangle = \frac{i}{8\pi^2} g^2 F_{\mu\nu}
                                    \tilde{F}_{\mu\nu}.
\eeq
in coordinate space, it is, therefore, necessary to regard the contribution
$\langle X(x)\rangle$ from the Wilson term (4) as a piece of the lattice
four-divergence of the axial vector current and write the axial Ward
identity in the Wilson model as
\beqa
\langle\frac{1}{2a}\{\left[\overline\psi_x\gamma_\mu\gamma_5
U_{x,\mu}\psi_{x+\mu}+\overline\psi_{x+\mu}U^\dagger_{x,\mu}\gamma_\mu\gamma_5
 \psi_x\right]-\left[x\rightarrow x-\mu\right]\}\rangle 
-\langle X(x)\rangle & & \nonumber \\
& &= 2i\;m\;\langle\overline\psi_x\gamma_5\psi_x\rangle,
\eeqa
\beq
\langle X(x)\rangle = \frac{i\;r}{a}\langle 2\overline\psi_x\gamma_5\psi_x
-\{\left[\overline\psi_x\gamma_5 U_{x,\mu}\psi_{x+\mu}+\overline\psi_{x+\mu}
U^\dagger_{x,\mu}\gamma_5\psi_x\right]+\left[x\rightarrow
x-\mu\right]\}\rangle.
\eeq
The contribution from the Wilson term thus plays a symmetric role with
respect to the vector and the axial vector Ward identities. In either case,
it is to be regarded as a piece of the four divergence of the respective
currents on the lattice.

It is remarkable that the specific form of the irrelevant term, in the
present case, the Wilson term (4), does not play any role in the derivation
of the anomaly (7) except that it must remove all the doublers, thereby
enabling the use of the Reisz theorem. Beyond this, we need 
gauge invariance to realize the tensor
structure (5), and locality so that the differential Ward identities between
the inverse propagator and photon vertices, e.g.,
\beq
V_\mu(p,p) = \frac{\partial}{\partial p_\mu} S^{-1}(p)
\eeq
and its generalizations \cite{KaSm81}, hold.

\vskip 0.5cm

\noindent
{\bf Chiral anomaly in a chirally symmetric model.}
We consider a lattice fermion action with a chirally symmetric irrelevant
term
\beqa
S_F &=& \sum_{x,\mu} \frac{1}{2a} \overline\psi_x \gamma_\mu
      [U_{x,\mu} \psi_{x+\mu} - U^\dagger_{x-\mu,\mu}\psi_{x-\mu}]
      +m\sum_x \overline\psi_x\psi_x \nonumber \\
   & &   + \sum_{x,\mu} \frac{1}{2a} \overline\psi_x \gamma_\mu \gamma_5
      [2\psi_x - U_{x,\mu}\psi_{x+\mu} - U^\dagger_{x-\mu,\mu}\psi_{x-\mu}],
\eeqa
proposed by us recently \cite{BanDe97}. The free fermion propagator is now
given by
\beq
\tilde S(p) = \left[\gamma_\mu \frac{\sin ap_\mu}{a} + i\gamma_\mu \gamma_5
              \frac{1-\cos ap_\mu}{a} + m \right]^{-1}. 
\eeq

The irrelevant term
\beq
i\gamma_\mu \gamma_5 \frac{1-\cos ap_\mu}{a},
\eeq
as in other local chirally symmetric models \cite{Others}, breaks hypercubic
and reflection symmetries. The present model (17), however, is hermitian. This
is an advantage in discussing reality properties in the continuum limit
\cite{Smit87}, and definitely in numerical simulations. Apart from the
$\gamma_\mu\gamma_5$, in the irrelevant term in
(17) one can recognize the scalar Wilson term. To
recover hypercubic symmetry in the correlation functions, it seems
necessary to average them over signs of the irrelevant term for each
direction \cite{Rothe}. It is, however, possible that 
for gauge-invariant physical amplitudes the continuum limit
itself takes care of the symmetry restoration as in Kogut-Susskind fermions.

The irrelevant term removes all the doublers from the spectrum
\cite{BanDe97}. This is evident from the absolute square of the inverse of
the fermion propagator (18),
\beq
\left[\tilde S(p)\tilde S(p)^\dagger\right]^{-1}
= \sum_\mu \left[ \frac{\sin^2 ap_\mu}{a^2} + \frac{(1-\cos ap_\mu)^2}{a^2}
\right ] + m^2  -  \frac{2}{a^2} \sum_{\mu\nu} \sigma_{\mu\nu} \gamma_5
\sin ap_\mu (1-\cos ap_\nu),
\eeq
whose {\it trace} vanishes in the chiral limit $m=0$ if and only if
$p_\mu=0$. The 1-loop vacuum polarization in QED has also been calculated
and shows expected behaviour in the continuum limit. The calculation of the
vacuum polarization will be reported elsewhere. 

The fermion action (17) leads to the axial Ward identity
\beq
\langle \frac{1}{a}\sum_\mu(\tilde J_{\mu 5}(x) - 
\tilde J_{\mu 5}(x-\mu) \rangle
= 2im \langle \overline\psi_x \gamma_5 \psi_x \rangle ,
\eeq
with the axial current given by,
\beq
\tilde J_{\mu 5}(x)= 
\frac{1}{2}\left[\overline\psi_x\gamma_\mu\gamma_5(1-\gamma_5)
U_{x,\mu}\psi_{x+\mu}+\overline\psi_{x+\mu}U^\dagger_{x,\mu}(1-\gamma_5)
\gamma_\mu\gamma_5\psi_x\right].
\eeq
The diagrams in Fig.1 represent, as in the Wilson case, the emission of two
photons arising from the right hand side of the Ward identity (21) (the
lower order diagrams can be shown to vanish).   
To construct the lattice amplitudes we need one-
and two-photon vertices
\beqa
\tilde V_\mu(p,q) &=& \gamma_\mu \cos \frac{a}{2} (p+q)_\mu
                      + i \gamma_\mu \gamma_5 \sin\frac{a}{2}(p+q)_\mu,
                      \nonumber \\
\tilde V_{\mu\nu}(p,q) &=& \delta_{\mu\nu} a \left[-\gamma_\mu
                           \sin\frac{a}{2}(p+q)_\mu + i\gamma_\mu\gamma_5
                           \cos\frac{a}{2}(p+q)_\mu \right] .
\eeqa 

The amplitudes of the diagrams (a) and (b), individually are not gauge
invariant. One finds instead
\beqa
\sum_\mu \frac{2}{a} \sin \frac{a}{2} p_\mu \tilde{D}^{(a+b)}_{\mu\nu}
&=& 4 g^2 m \int_l Tr [i\gamma_5 \tilde{S}(l)\{ \tilde{V}_\nu(l,l+q)
    -\tilde{V}_\nu(l+p,l+p+q)\}\tilde{S}(l+p+q)]\nonumber \\
&=& -8g^2 m \sum_\mu \int_l Tr [i\gamma_5 \tilde{S}(l)\tilde{V}_{\mu\nu}
    (l,l+p+q) \tilde{S}(l+p+q)] \sin\frac{a}{2}p_\mu \nonumber \\
&=& -\sum_\mu \frac{2}{a} \sin\frac{a}{2}p_\mu \tilde{D}^{(c)}_{\mu\nu}.
\eeqa 

Thus, in the present case only the sum of amplitudes of all the three
diagrams (a), (b) and (c) is gauge-invariant and has the structure
\beq
\tilde{D}^{(a+b+c)}_{\mu\nu} \propto \epsilon_{\mu\nu\alpha\beta}p_\alpha
q_\beta.
\eeq

The amplitude for the diagram (c) is a function of $(p+q)$, and thus
symmetric in $p$ and $q$. Therefore, it cannot contribute to the gauge
invariant structure (25). The leading terms in the amplitudes for the
diagrams (a) and (b) are of degree $1$ by Reisz power counting, as before,
but in the present case they do not contribute because of
vanishing trace of odd number of $\gamma$-matrices. The coefficient of
$p_\alpha q_\beta$ in the gauge-invariant structure (25) has, therefore, an
effective degree $-2$ as in the case of Wilson model. Reisz theorem allows, as
before, to take the continuum limit $a\rightarrow 0$ within the lattice
integral. Due to this, the dependence of the vertices $\tilde V_\mu(l-p,l)$
and $\tilde V_\nu(l,l+q)$ in
\beq
\tilde D^{(a)}_{\mu\nu}=-2g^2m\int_l Tr \left[i\gamma_5 \tilde S(l-p) 
\tilde V_\mu(l-p,l)\tilde S(l)\tilde V_\nu(l,l+q) \tilde S(l+q)\right]
\eeq
on the external momenta $p$ and $q$ respectively can be ignored. This can be
easily verified by taking the derivative of $\tilde V_\mu$ with respect to
$p_\mu$. The resulting contribution vanishes in the continuum limit. In
extracting the gauge invariant structure (25), the dependence of only the
propagators $\tilde S(l-p)$ and $\tilde S(l+q)$ on external momenta is
relevant.

The gauge invariant contributions from the diagrams (a) and (b) thus
coincide with the same in the Wilson model and ABJ anomaly (7) is reproduced in
the present model (17),
\beq
\lim_{a\rightarrow 0} \tilde D^{(a+b)}_\mu = \frac{i}{2\pi^2}g^2
\epsilon_{\mu\nu\alpha\beta}p_\alpha q_\beta .
\eeq


\vskip 0.5cm
{\bf Conclusions.}
At finite lattice spacing $a$, there are no anomalous Ward identities on the
lattice. Anomalies, if any, appear only in the continuum limit through
correspondences assumed between lattice operators and their continuum
counterparts \cite{Fuji}. It is natural that the contribution from the
irrelevant term, a lattice artifact, is identified as the generator of the
anomaly in axial Ward identity \cite{KaSm81,Kerler83}.This
identification, it should be emphasized, is at best a plausible assumption.
Indeed, in the case of lattice Ward identity for the vector current, the
contribution from the irrelevant term is included in the definition of the
four-divergence of the vector current in the continuum \cite{KaSm81}. In the
present approach, however, we have identified the contribution of the
physical mass of the fermion as the generator of the ABJ anomaly. The ABJ
anomaly in this approach consists  in the difference in the continuum limit
of the four-divergence of the axial current in a gauge theory with massless
($m=0$) fermion and that obtained in the chiral limit $m\rightarrow 0$
starting with a massive fermion. This is how the results of Lee and
Nauenberg \cite{LN64} and of Dolgov and Zakharov \cite{DZ71} in continuum
QED are realized on lattice.


\begin{thebibliography}{99}

\bibitem{NN81} H.B. Nielsen and M. Ninomiya, Nucl. Phys. B185 (1981) 541. 

\bibitem{Pelli87} A. Pelissetto, Ann. Phys. (N.Y.) 182 (1988) 177; M.
Pernicci, Phys. Lett. B346 (1995) 99. 

\bibitem{Wilson77} K.G. Wilson, in {\it New Phenomena in Subnuclear Physics}
                   (Erice, 1975), ed. A. Zichichi (Plenum Press, New York, 1977),
                   p.69.

\bibitem{KaSm81} L.H. Karsten and J. Smit, Nucl. Phys. B127 (1981), 103.  

\bibitem{Kerler83} W. Kerler, Phys. Rev. D 23 (1981) 2384; E. Seiler and 
                   I.O. Stamatescu, Phys. Rev. D 25 (1982) 2177; T. Hattori
and H. Watanabe, hep-lat/9708007.  

\bibitem{LN64} T.D. Lee and M. Nauenberg, Phys. Rev. B 133 (1964) 1549.

\bibitem{DZ71} A.D. Dolgov and V.I. Zakharov, Nucl. Phys. B27 (1971) 525;
K. Huang, in {\it Quarks, Leptons and Gauge Fields}(World
Scientific, 1982), p219. 

\bibitem{Reisz} T. Reisz, Comm. Math. Phys. 116 (1988) 81; Comm. Math. Phys.
                116 (1988) 573.

\bibitem{Luescher} M. Luescher, in {\it Fields, strings and critical
Phenomena} (Les Houches, 1988), 
eds. E. Brezin and J. Zinn-Justin, (North Holland, Amsterdam,
1989), p451.

\bibitem{Fuji}  K. Fujikawa, Z. Phy. C25 (1984) 179.

\bibitem{BanDe97} H. Banerjee and Asit K. De, Nucl. Phys. B (Proc. Suppl.)
53 (1997) 641.

\bibitem{Others} L. Jacobs, Phys. Rev. Lett. 51 (1983) 172; D. Weingarten
and B. Velikson, Nucl. Phys. B270 (1986) 10; J.L. Alonso and J.L. Cortes,
Phys. Lett. B187 (1987) 146; I.O. Stamatescu and T.T. Wu, Nucl. Phys. B
(Proc. Suppl.) 42 (1995) 838.

\bibitem{Smit87} J. Smit, Nucl. Phys B (Proc. Suppl.) 4 (1988) 451.

\bibitem{Rothe}  N. Sadooghi and H.J. Rothe, Phys. Rev. D55 (1997) 6749.

\end{thebibliography}
\end{document}